\documentclass[pdflatex,sn-apa]{sn-jnl}

\usepackage{url}
\usepackage[strings]{underscore}
\usepackage{graphicx}%
\usepackage{multirow}%
\usepackage{amsmath,amssymb,amsfonts}%
\usepackage{amsthm}%
\usepackage{mathrsfs}%
\usepackage[title]{appendix}%
\usepackage{xcolor}%
\usepackage{textcomp}%
\usepackage{manyfoot}%
\usepackage{booktabs}%
\usepackage{algorithm}%
\usepackage{algorithmicx}%
\usepackage{algpseudocode}%
\usepackage{listings}%
\usepackage{tabularray}
\usepackage{pdflscape}
\usepackage[sectionbib]{bibunits}

\usepackage[nolist]{acronym}

\usepackage{enumitem}
\usepackage{array}
\usepackage{longtable}
\usepackage{cleveref}

\usepackage{titlesec}
\titleformat{\paragraph}[runin]{\normalfont\bfseries}{\thesection.}{.5em}{}
\titlespacing{\paragraph}{\parindent}{1.5ex plus .1ex minus .2ex}{2ex}

\begin{document}
\title{Building Symbiotic Artificial Intelligence: Reviewing the AI Act for a Human-Centred, Principle-Based Framework}

\author*[1]{\fnm{Miriana} \sur{Calvano}}\email{miriana.calvano@uniba.it}
\author[1,2]{\fnm{Antonio} \sur{Curci}}\email{antonio.curci@uniba.it}
\author[1]{\fnm{Giuseppe} \sur{Desolda}}\email{giuseppe.desolda@uniba.it}
\author[1]{\fnm{Andrea} \sur{Esposito}}\email{andrea.esposito@uniba.it}
\author[1]{\fnm{Rosa} \sur{Lanzilotti}}\email{rosa.lanzilotti@uniba.it}
\author[1]{\fnm{Antonio} \sur{Piccinno}}\email{antonio.piccinno@uniba.it}

\affil[1]{\orgdiv{Department of Computer Science}, \orgname{University of Bari Aldo Moro}, \orgaddress{\street{Via E. Orabona 4}, \city{Bari}, \postcode{70125}, \country{Italy}}}

\affil[2]{\orgdiv{Department of Computer Science}, \orgname{University of Pisa}, \orgaddress{\street{Largo B. Pontecorvo, 3}, \city{Pisa}, \postcode{56127}, \country{Italy}}}

\abstract{%
Artificial Intelligence (AI) spreads quickly as new technologies and services take over modern society. The need to regulate AI design, development, and use is strictly necessary to avoid unethical and potentially dangerous consequences to humans. The European Union (EU) has released a new legal framework, the AI Act, to regulate AI by undertaking a risk-based approach to safeguard humans during interaction. At the same time, researchers offer a new perspective on AI systems, commonly known as Human-Centred AI (HCAI), highlighting the need for a human-centred approach to their design. In this context, Symbiotic AI (a subtype of HCAI) promises to enhance human capabilities through a deeper and continuous collaboration between human intelligence and AI. This article presents the results of a Systematic Literature Review (SLR) that aims to identify principles that characterise the design and development of Symbiotic AI systems while considering humans as the core of the process. Through content analysis, we elicit four principles that must be applied to create Human-Centred AI systems that can establish a symbiotic relationship with humans. In addition, current trends and challenges are presented to indicate open questions that may guide future research for the development of SAI systems that comply with the AI Act.
}

\maketitle

\begin{acronym}[HCAI]
    \acro{AI}{Artificial Intelligence}
    \acro{UI}{User Interface}
    \acro{HCAI}{Human-Centred Artificial Intelligence}
    \acro{GDPR}{General Data Protection Regulation}
    \acro{SLR}{Systematic Literature Review}
    \acro{HCI}{Human-Computer Interaction}
    \acro{HCD}{Human-Centred Design}
    \acro{EU}{European Union}
    \acro{SAI}{Symbiotic Artificial Intelligence}
    \acro{LLM}{Large Language Model}
    \acro{ML}{Machine Learning}
    \acro{DL}{Deep Learning}
    \acro{HLEG}{High Level Expert Group}
\end{acronym}

\section{Introduction}\label{introduction}
In recent years, society has witnessed a significant surge in interest and investment in \ac{AI}, primarily driven by advancements in \ac{ML} and \ac{DL}. These technologies have become transformative forces, enabling groundbreaking innovations and offering new services across a wide spectrum of domains, from healthcare and finance to transportation and entertainment. Such progress underscores the growing role of \ac{AI} in shaping modern society and highlights the urgency of understanding its applications, implications, and potential.
Despite its advancements, in fact, \ac{AI} raises significant ethical, legal, and human rights concerns, ranging from fear of discrimination \citep{Stahl2023Unfair} to deskilling \citep{Sambasivan2022Deskilling}.

The burgeoning pervasiveness of \ac{AI} in daily-use systems has highlighted several major flaws in current AI techniques. Among these, biases and lack of explainability greatly endanger the users of \ac{AI} models \citep{Shneiderman2020Bridging}, who may be treated unfairly or exposed to life-threatening risks \citep{NationalTransportationSafetyBoard2017Collision}.

These concerns have prompted academics and global legislative bodies to take action to ensure the responsible development of \ac{AI}. In the academic sphere, a new perspective known as \ac{HCAI} \citep{Xu2019HumanCentered,Shneiderman2022HumanCentered} is reshaping the research landscape.
Located at the intersection between \ac{HCI} and \ac{AI}, \ac{HCAI} promises to provide a direction to design AI systems that are safe, reliable, and trustworthy \citep{Shneiderman2020HumanCentered}. More precisely, \ac{HCAI} aims to design, develop, and evaluate \ac{AI} systems involving end-users in the process to increase their performances and satisfaction in performing specified tasks \citep{Desolda2024Humancentered}. Therefore, \ac{HCAI} systems aim to be \emph{usable} and \emph{useful} for specified users to reach their specified goals in their context of use, while being reliable, safe to use, and trustworthy \citep{Desolda2024Humancentered}.
By adopting a human-centred design approach, \ac{HCAI} may foster a \emph{symbiotic relationship} between humans and \ac{AI}. Therefore, \ac{SAI} systems are a specific type of \ac{HCAI} systems that allow for continuing and deeper collaborations between human intelligence and \ac{AI}, aiming at their mutual augmentations without hampering human autonomy or \ac{AI}'s performance \citep{Grigsby2018Artificial, Desolda2024Humancentered,Szafran2024Human}. However, this may not suffice to ease ethical concerns.

Simultaneously, governments are beginning to draft and implement new legislative measures. A notable example is the \ac{EU}'s \ac{AI} Act, which represents a pioneering effort in regulating \ac{AI} \citep{EuropeanParliament2024Regulation}.
The \ac{AI} Act is a legal framework concerning \ac{AI} systems' design, development, deployment, and use, aiming at ensuring their proper employment while minimising risks to human well-being and society. It leverages a human-centric and risk-based approach \citep{Mokander2022Conformity}, considering humans in all their dimensions and not as mere users, promoting fundamental rights and categorising \ac{AI}-based systems into four distinct risk levels: \emph{Low-Minimal}, \emph{Limited}, \emph{High}, and \emph{Unacceptable}. This classification is based on the \ac{AI} system's purpose and its potential impact on human well-being, with stricter rules applied as the level of risk increases \citep{EuropeanParliament2024Regulation}.

This work aims to identify the principles that can guide the design of \ac{SAI} systems to comply with the current legal landscape considering humans as the core of the development process \citep{ISO20199241210}. With this goal in mind, we performed a \ac{SLR}, employing content analysis to understand what researchers suggest about the relationship between \ac{SAI} and the \ac{AI} Act.
The main contribution of this study is the proposal of principles to guide stakeholders in the design of \ac{SAI} systems, i.e., of \ac{HCAI} systems that favour a \emph{symbiotic} relationship with humans. It is finally put in the context of the relevant literature, comparing it with other principled-based approaches to \ac{SAI} and \ac{HCAI}. The goal of this study is thus better formulated in the following research question:

\begin{center}
\emph{RQ: What principles can guide the design of \ac{SAI} systems, while promoting a symbiotic human-\ac{AI} relationship, and adhering to the European \ac{AI} Act?}
\end{center}

The article is structured as follows: \Cref{methodology} describes the methodology adopted to conduct the \ac{SLR}; \Cref{results} presents the results obtained by the \ac{SLR} (i.e., principles and properties that characterize the human-\ac{AI} symbiosis); \Cref{discussion} discusses the analysis of the literature presenting the challenges, trends, and limitations faced while conducting the \ac{SLR}.

\section{Methodology}\label{methodology}

The systematic literature review was carried out following Kitchenham's procedure \citep{Kitchenham2004Procedures}. The procedure reinforces the key steps of the review to ensure transparency, minimise bias, and contribute to the advancement of knowledge by defining clear research questions and protocol compliant with the objective.

The keywords of the \ac{SLR} were defined as follows. Being the core of our research, the keyword \emph{artificial intelligence act} was necessary to identify the context of the investigation, while \emph{human-centric artificial intelligence} was identified as the \ac{AI} Act takes on a human-centric approach; we decided not to include ``human-centred'' because the \ac{AI} Act views it as a different approach and not as a synonym \citep{EuropeanParliament2024Regulation}. The keyword \emph{intelligent systems} was used to encompass all the systems that exhibit an intelligent behaviour and was used as a synonym of \ac{AI}. Finally, \emph{symbiotic artificial intelligence} was a crucial keyword to explore the field of \ac{AI}-based systems that establish a symbiotic relationship with humans. Starting from these keywords and based on the research objectives, the following queries were built:

\begin{enumerate}[label={(Q\arabic*)}, align=left]
\item european AND (``artificial intelligence'' OR ai) AND act
\item european AND (``artificial intelligence'' OR ai) AND act AND (human OR human-centric OR human-centred OR human-centered)
\item (``human centric artificial intelligence'' OR ``human centric AI'') AND (``artificial intelligence act'' OR ``ai act'')
\item (``symbiotic artificial intelligence'' OR ``Symbiotic AI'') AND (``artificial intelligence act'' OR ``ai act'')
\item ``Intelligent Systems'' AND (``artificial intelligence act'' OR ``ai act'')
\end{enumerate}

\subsection{Inclusion and Exclusion Criteria}\label{inclusion-and-exclusion-criteria}

The inclusion and exclusion criteria were defined based on the research objectives. They are listed and described below.

\begin{itemize}
\item \emph{Year}: the period from 2022 to 2024 was considered because the \ac{AI} Act was still an early draft before 2022 and underwent substantial changes throughout small periods.
\item \emph{Topic}: we included papers that revolve around the \ac{AI} Act and its implications in different areas of science and society, but documents in which the legal framework was merely mentioned or slightly addressed were excluded.
\item \emph{Peer reviewed}: for articles appearing in journals, we included those ranked Q1, Q2, and Q3 in Scimago; for conference articles, we included A, B, and C conferences on the Core Conference Ranking. Conference articles ranked as regional can be considered if they can significantly impact the review results.
\item \emph{Document type}: the review focused on single manuscripts and papers, excluding from the queries’ outputs the entire proceedings of conferences or books.
\item \emph{Language}: each paper not written in English was excluded.
\end{itemize}

\subsection{Coding and Classification}\label{coding-and-classification}

The papers were classified using a mixed approach of data coding: i.e., \emph{a-priori} and \emph{in-vivo}. \emph{A-priori} coding consists of the categorisation of papers with codes that are established before the classification process. In contrast, \emph{in-vivo} coding involves the definition of the codes as the papers are read and analysed \citep{Lazar2017Research}. First, articles were classified according to the a-priori defined codes to guarantee coherence in the analysis. Then, in-vivo coding was adopted to make the methodology feasible and integrate new and emerging concepts. During this phase, the name of each code, along with its sub-codes, was carefully and iteratively refined with respect to the topics that characterize the specific group of articles.

The name of each principle and its dimensions correspond to the name of codes and sub-codes defined during the process. They were collectively chosen by the researchers who took part in the \ac{SLR}, reflecting terms that are recurrent in the set of papers and those used in the regulation.
\subsection{Conducting the SLR}\label{conducting-the-slr}

\begin{figure}[th!]
    \centering
    \includegraphics[width=.8\linewidth]{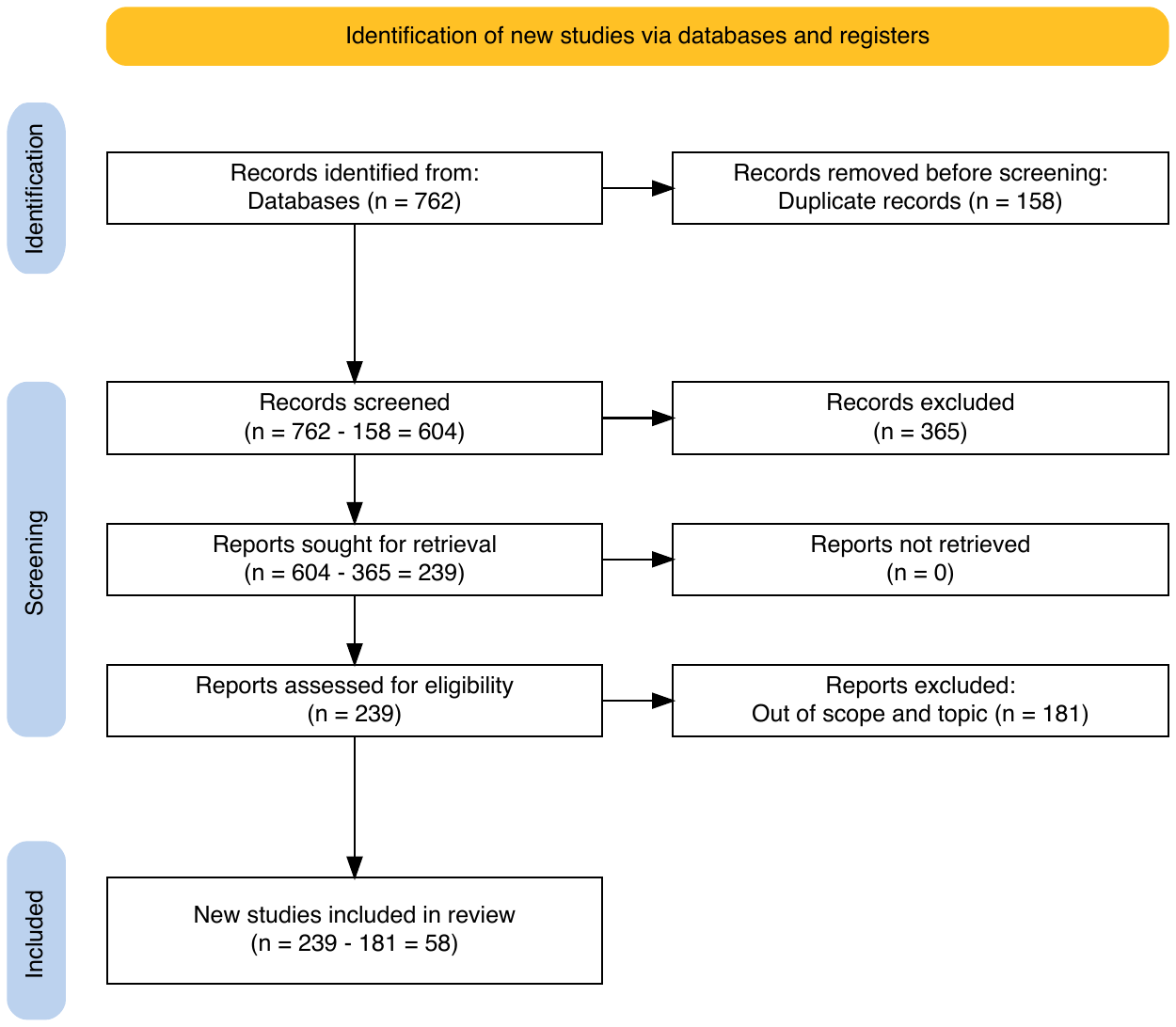}
    \caption{PRISMA diagram of the literature review}
    \label{fig:prisma}
\end{figure}

The first step of the execution was searching for academic articles about the topic by running the five queries on the Scopus digital library because it is comprehensive and includes the most relevant and accredited digital libraries, including journal and conference manuscripts. Each query provided the following results: Q1: 530, Q2: 223, Q3: 4, Q4: 1, and Q5: 4. It emerged that Q1 and Q2 returned a higher number of papers obtained since they are broader and more exploratory than the others.

The resulting set of papers from the queries underwent a selection process respecting both inclusion and exclusion criteria.

From a total of 762 papers, after a check for duplicates (i.e., 158 files), 604 publications were obtained. A more in-depth step was taken, analysing titles and keywords, which resulted in the removal of 365 papers since they did not match the inclusion criteria.
In the end, each publication was further analysed by reviewing the abstract, the introduction, and the conclusions, obtaining the final set of papers. The final set used for the literature review contains 58 papers. \Cref{fig:prisma} shows the PRISMA diagram of the whole process \citep{Page2021PRISMA}.

The identified 58 publications are presented in \Cref{tab:map-articles-principles}, showing the mapping with the principles and their dimensions. \Cref{tab:map-principles-properties} provides an overview of which article discusses a property in the context of each principle. The framework represented in \Cref{fig:framework} is elicited through a qualitative analysis of the data represented in \Cref{tab:map-articles-principles,tab:map-principles-properties}.

\section{Results: A Principled Human-Centred AI Framework to Build SAI}\label{results}

\begin{figure}[t!]
    \centering
    \includegraphics[width=\linewidth]{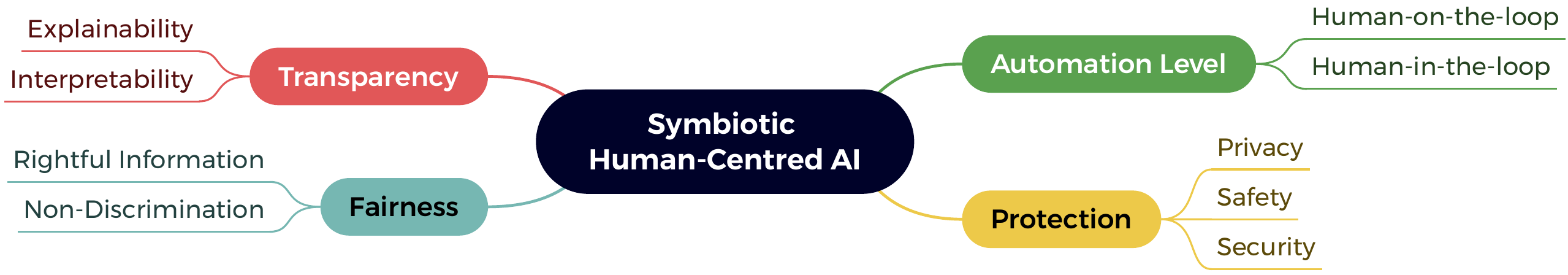}
    \caption{Principles for Symbiotic Human-Centred \ac{AI} and their dimensions}
    \label{fig:principles}
\end{figure}

The four principles, represented in \Cref{fig:principles}, and their relationship with the additional three properties are detailed in the following sections, discussing how they can support the creation of \ac{SAI} systems, i.e., \ac{HCAI} systems that foster a symbiotic relationship with humans \citep{Desolda2024Humancentered}.

The principles emerged during the coding phase. In a first iteration of the \ac{SLR}, the report by \cite{Fjeld2020Principled} was taken as the a-priori coding scheme since it presents a principle-based framework for \ac{AI} systems before the release of the \ac{AI} Act. In order to determine how the regulation impacted the creation of such systems, the coding scheme was modified during later iterations through the in-vivo technique. After the analysis of the articles and the definition of the final codes, additional intrinsic and transversal aspects emerged, which were identified as properties.

\subsection{Principles for Human-AI Symbiosis}\label{principles-for-human-ai-symbiosis}
Defining principles to guide the design and development of \ac{AI} systems through a human-centred approach is crucial to establish a symbiotic relationship between the two parties \citep{Desolda2024Humancentered}. It is important to guarantee that \ac{AI} systems adhere to the highest standards of ethical conduct. Thus, designers and developers must ensure that the implemented solutions comply with laws and regulations safeguarding human rights and European values, making these systems safer for end-users \citep{Helberger2023ChatGPT,Worsdorfer2023Mitigating}. To this end, providers must register high-risk \ac{AI} models in an \ac{EU} database managed by the European Commission to enhance public transparency and enable oversight by authorities \citep{Pavlidis2024Unlocking}.

This section explores the four principles identified through the \ac{SLR}, describing their dimensions and characteristics.

\subsubsection{Transparency}\label{principle-1-transparency}

\emph{Transparency} can guarantee that \ac{AI} systems are effectively overseen by humans and allow intervention when potential harm occurs \citep{Pavlidis2024Unlocking}. Transparency ensures that critical information about how the \ac{AI} model was trained and structured is available to humans \citep{Pavlidis2024Unlocking}, since stakeholders must understand how \ac{AI} models function and the reasoning behind their decisions to be able to intervene \citep{Pavlidis2024Unlocking}. Being a multi-faceted property that concerns \ac{AI} models, their components, and algorithms, \emph{Transparency} serves as an \emph{umbrella} term to the functional understanding of the model and the rationale behind its operations \citep{Gyevnar2023Bridging,Urquhart2022Legal}. It is characterized by two dimensions: \emph{Explainability} and \emph{Interpretability} \citep{HighLevelExpertGrouponArtificialIntelligence2020Assessment}.

\paragraph{Explainability}\label{explainability}
Explainability aims to provide explanations concerning \ac{AI} systems' operations, ensuring that even when humans cannot understand how an \ac{AI} system provides an output, they can at least receive information about why it was produced \citep{Siegel2024Media}. In other words, this aspect contributes to the degree to which an \ac{AI} system is open and observable to humans \citep{Pavlidis2024Unlocking,Urquhart2022Legal}. When interacting with \ac{AI} systems, humans should have the \emph{right to explanations} to make informed decisions, supported by \ac{AI} and not replaced by it \citep{Malgieri2024Licensing,Wagner2024Navigating}, while being able to oversee its processes \citep{Tartaro2023European}.

\paragraph{Interpretability}\label{interpretability}
An \ac{AI} system can be considered interpretable if it can be correctly understood by an individual who can assign meanings to outputs \citep{HighLevelExpertGrouponArtificialIntelligence2020Assessment}. This means that the person interacting with \ac{AI} can understand its functionality, purpose, or impact on the context of use. Therefore, the process of interpretation involves mapping an abstract concept, such as a predicted class or category, into a domain that is within the grasp of human understanding \citep{Pavlidis2024Unlocking}.

\subsubsection{Fairness}\label{principle-2-fairness}
In \ac{AI} systems, Fairness reflects the concepts of equality and inclusiveness to avoid biases and discriminatory behaviors safeguarding fundamental human rights and values \citep{HighLevelExpertGrouponArtificialIntelligence2020Assessment}. An approach that considers humans as a whole is essential for ensuring that \ac{AI} can contribute positively to society and protect them against potential harms \citep{Anamaria2023Artificial}. Fairness is characterized by two dimensions: \textit{Rightful Information} and \textit{Non-Discrimination}.

\paragraph{Rightful Information}\label{rightful-information}
\ac{AI} systems must disseminate accurate and reliable information to minimise the risk of incorrect or incomplete knowledge and to mitigate the risk of manipulation and persuasion of humans \citep{Herbosch2024Fraud,Sposini2024Neuromarketing}. Providing accurate and complete information is essential to explain \ac{AI} decisions to safeguard individual rights and freedom, foster a sense of trust and understanding among humans, and enhance the ethical use of \ac{AI} systems \citep{Worsdorfer2023Mitigating}.

\paragraph{Non-Discrimination}\label{non-discrimination}
Although \ac{AI} is commonly used to boost productivity through automation, it is important to ensure that models are trained in accordance with ethical and societal norms that minimize discriminatory behaviors, avoiding that individuals are treated differently or unequally without any justifiable reason \citep{Biewer2024Software}. Thus, the whole pipeline of creating \ac{AI}-based systems must be monitored and checked since biases can rise unintentionally during the early training phases of models \citep{Nikiforov2024Groups,Pavlidis2024Unlocking}.

\subsubsection{Automation Level}\label{principle-3-automation-level}
As \ac{AI} becomes increasingly integrated into countless aspects of human life, studying the appropriate balance between automation and human control in human-\ac{AI} interactions is necessary \citep{Enqvist2023Human}. Although there are contexts of use in which humans need or wish for fully automated systems in which their control is not necessary, it is important to address the ethical and legal consequences of undesired events caused by \ac{AI} systems' outputs. This implies that automation can be considered as a spectrum and not as a binary feature \citep{Esposito2024Fine}. Automation Level is characterized by two dimensions: \textit{Human-on-the-loop} and \textit{Human-in-the-loop}.


\paragraph{Human-on-the-loop}\label{human-on-the-loop}
Providing humans with appropriate oversight when interacting with \ac{AI} systems is necessary to enable them to check, monitor, and supervise the system's behaviour. Oversight is a precondition to allow human intervention, which guarantees \ac{AI}-assisted decisions rather than \ac{AI}-driven decisions. This can avoid irreversible consequences and minimise risk and biased outputs while safeguarding human rights \citep{CoveloDeAbreu2024Artificial}. Human oversight is strictly related to \emph{Transparency}, as \ac{AI} models must provide effective explanations to users, ensuring that users can effectively interpret outputs in order to properly modify their behaviour, if necessary \citep{Hupont2022Landscape}.

\paragraph{Human-in-the-loop}\label{human-in-the-loop}
Designing \ac{AI} systems emphasising human control is useful in situations where humans need to actively participate in the decision-making process. In this context, the concept of \textit{Controllable \ac{AI}} is introduced, which reinforces the importance of human control to detect malfunctions and recover from dangerous situations; controlling the behavior of \ac{AI} means influencing its output and processes in accordance with the context of use and human expertise for a more safe and reliable interaction \citep{Kieseberg2023Controllable}. Some of the conditions must be in place for humans to take control of the interaction properly: appropriate algorithmic transparency of \ac{AI} models and high levels of feedback and affordance in \acp{UI} are needed since communication is key in any kind of relationship \citep{Siegel2024Media}.

\subsubsection{Protection}\label{principle-4-protection}

The human-centric approach undertaken by the \ac{AI} Act aims at ensuring that users are safeguarded against harm, threats, or intrusion. This principle is strongly intertwined with the legal requirements set by governmental norms and rules that designers, developers, and deployers must comply with to protect users from unsafe behavior. There is a need to create secure and resilient \ac{AI} systems that can preserve users' privacy. In this regard, the \ac{AI} Act recalls the \ac{GDPR}, which emphasized the integration of privacy and data protection into the design and development of systems \citep{EuropeanParliament2024Regulation}. Protection refers to a large spectrum of aspects, laying its ground in cybersecurity, and being characterized by three dimensions: \emph{Privacy}, \emph{Safety}, and \emph{Security}.

\paragraph{Privacy}\label{privacy}
An \ac{AI} system that fosters users' privacy can safeguard individuals' sensitive data from improper access, theft, or loss \citep{Fjeld2020Principled}. This principle transcends from the specific case of \ac{AI} systems, as applying the proper techniques to preserve data and protect individuals' identities impacts multiple system components and requires developers to inform end-users about how their data is being handled \citep{Pavlidis2024Unlocking}. The correct application of data preservation techniques and protective measures can have far-reaching impacts on multiple system components.

\paragraph{Safety}\label{safety}
A safe system is designed to fulfill its intended function without causing harm to living beings or the environment \citep{Fjeld2020Principled}. This concept is linked to the well-being and welfare of humans affected by \ac{AI} and is also connected to the system's level of automation. The objective is to mitigate risk and prevent accidents by removing barriers to error reporting and fostering a collaborative and communicative environment. This ensures that end-users are always informed about potentially harmful practices that could threaten their rights \citep{Neuwirth2023Prohibited}.

\paragraph{Security}\label{security}
To protect individuals, systems must be \emph{secure by design}, implying that they must incorporate solutions that allow management, monitoring, and recovery from external threats \citep{Nikiforov2024Groups}. This dimension is relevant to all types of systems, regardless of \ac{AI} features, since security implies the preservation of the \emph{CIA} triad (Confidentiality, Integrity, and Availability), meaning that strong prevention and recovery measures must be implemented. \ac{AI} systems, especially those categorized as \textit{high-risk} by the \ac{AI} Act, should be designed to be resilient against attacks and consistently perform securely throughout their lifecycle since they deal with extremely private and sensitive information about their users \citep{Siegel2024Media}.

\subsection{Properties of Human-AI Symbiosis}\label{properties-of-human-centred-ai}
\begin{figure}[t]
    \centering
    \includegraphics[width=.85\linewidth]{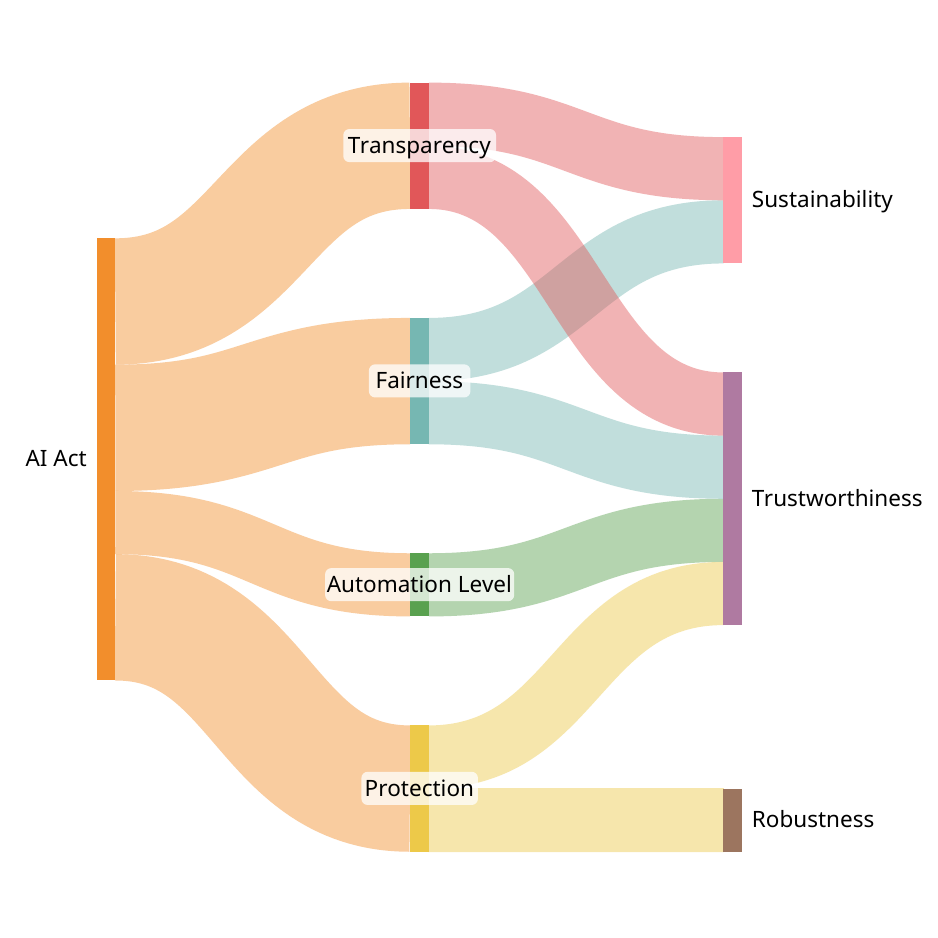}
    \caption{The AI Act-compliant framework of principled-based Symbiotic Human-Centred \ac{AI}}
    \label{fig:framework}
\end{figure}

Three properties emerged from the \ac{SLR}, which underlie the principles that were identified and that must characterise \ac{AI}-based systems that are designed and developed compliantly with the \ac{AI} Act. After the coding phase, an additional meta-analysis was performed using hierarchical clustering, and after the initial step of identifying the first set of clusters, a hierarchical grouping was brought to the identification of the principles described in the previous section. A further grouping step allowed us to identify the overall properties of a truly human-\ac{AI} symbiotic relationship: \emph{Trustworthiness}, \emph{Robustness}, and \emph{Sustainability}. This section details these properties, highlighting their relationship with the four principles described in \Cref{results}. It is highlighted that these properties are not a sufficient condition for AI systems to be symbiotic, but they emerged as crucial components of the design phase from this \ac{SLR}, influenced by its goals. \Cref{fig:framework} shows the connection among the identified properties and principles.

\subsubsection{Trustworthiness}\label{trustworthiness}

\emph{Trustworthiness} is a critical, frequently-discussed topic in the context of the human-centric approach of the \ac{AI} Act. It is essential for high-risk systems because it is necessary for reliable, safe, and positive interactions \citep{Stettinger2024Trustworthiness}. It can be fostered by implementing appropriate transparency techniques in \ac{AI} models in order to make users aware of the motivations that lie behind outputs and responses \citep{Laux2024Trustworthy}. This property is still a subject of debate within the research community and the legal landscape. For example, in the context of \ac{HCAI}, a trustworthy system is one that deserves human trust, implying that it must align with users' needs, preferences, and cognitive models to achieve successful interactions \citep{Shneiderman2022HumanCentered}. On the other hand, the \ac{AI} \ac{HLEG} identifies trustworthiness as the umbrella property to ensure a human-centric approach to \ac{AI} \citep{High-LevelExpertGrouponArtificialIntelligence2019Ethics}. Nevertheless, our analysis showed that trustworthiness is an important property, which covers the four principles, but it is not a sufficient condition for the establishment of a symbiotic relationship \citep{Desolda2024Humancentered}. It emerges that an \ac{AI} system is trustworthy if it enables humans to properly oversee and/or control its performance with an appropriate automation level while exhibiting fair behavior and respecting humans in all their dimensions.

\subsubsection{Robustness}\label{robustness}

Referring to the \ac{AI} Act, the \emph{Robustness} of \ac{AI} has been defined as their ability to perform reliably and effectively under various conditions, including unexpected or challenging ones. Robustness is mentioned in Article 15 of the law, which is titled ``Accuracy, Robustness and Cybersecurity'' in the context of ensuring that high-risk systems feature fail-safe plans and technical redundancy solutions \citep{EuropeanParliament2024Regulation}. The motivations behind this lie in the fact that a robust \ac{AI} system exhibits a safe, resilient, and reliable behaviour, which fosters trust in humans, thus being one of the factors that contribute to a symbiotic human-centred relationship \citep{Laux2024Trustworthy,Tallberg2024AI}.


\subsubsection{Sustainability}\label{sustainability}

\emph{Sustainability} is a factor that deployers should consider when creating systems to align with the \ac{EU}'s goals for the near future. As environmental concerns are substantially increasing worldwide and given the \ac{EU}'s efforts in trying to reduce emissions and energy consumption, a sustainable approach is necessary to minimize the environmental impact of \ac{AI} and to create long-lasting products that uphold ethical standards \citep{Pagallo2022Environmental,Siegel2024Media}. For example, reducing the complexity of Neural Networks can benefit \ac{AI} systems, as it reduces the required computational power in the training phase while increasing transparency \citep{Schwartz2020Green}. Although there are best practices that designers and developers should follow and requirements set by the law, it is highlighted that there is a lack of standardized methodologies for evaluating the impact of \ac{AI} systems in terms of sustainability \citep{Hacker2023European}. As the \ac{AI} Act was refined, the \ac{EU} has made steps forward in this direction by releasing compliance-checkers and user-friendly explorers of the law, but they remain auto-assessment tools that do not foster compliance by design. \ac{AI} systems must comply with the law and satisfy the requirements in terms of sustainability, being reliable and resilient regardless of the environmental and contextual changes that they undergo.

In summary, through content analysis performed on 58 articles, four principles were identified as requirements for designing \ac{AI} Act-compliant \ac{SAI} systems: \emph{Transparency, Fairness, Automation Level, and Protection}. Through further analysis, three additional properties were identified: \emph{Trustworthiness}, \emph{Robustness}, and \emph{Sustainability}.

\section{Discussion}\label{discussion}

This section discusses the analysis of the literature concerning the challenges, trends and limitations faced while conducting the \ac{SLR}. The release of the \ac{AI} Act has shifted the focus towards a more human-centric approach even in the literature belonging to the more technical side of Computer Science.
Individuals are not considered as mere users, but as human beings in all their dimensions, which must be included in the process of creating ethical \ac{AI} systems \citep{Worsdorfer2023Mitigating}. This aspect must not undermine the performance and flexibility of such systems, providing humans with context-aware solutions that are able to support and not replace them \citep{Carnevale2024Humancentred}. Nevertheless, little indication is provided regarding the design patterns that can be employed to implement human control and oversight mechanisms in \ac{AI} systems.

\subsection{Grounding the Framework in the Literature}\label{grounding-the-framework-in-the-literature}

\begin{figure}[t]
    \centering
    \includegraphics[width=\linewidth]{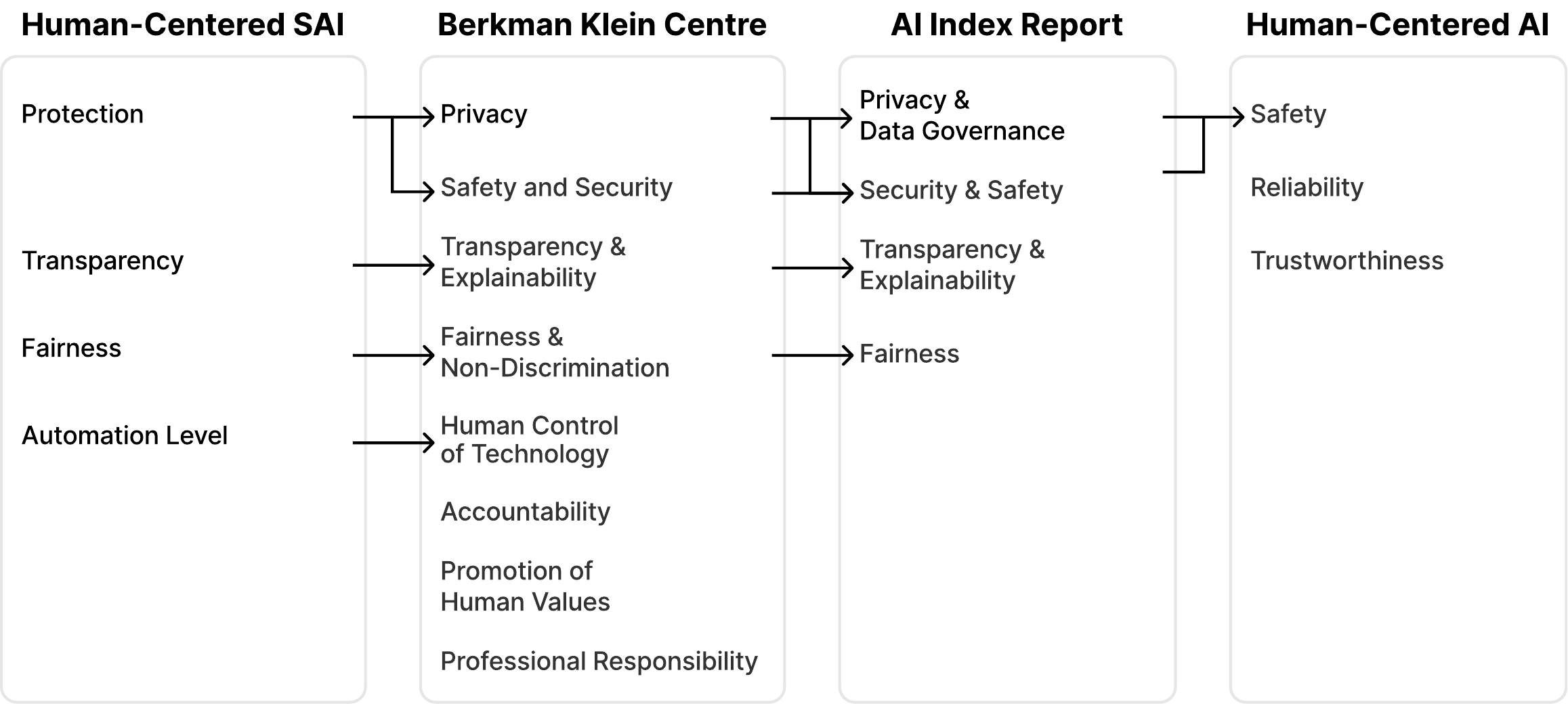}
    \caption{Mapping between the different sets of principles provided by the reference guidelines}
    \label{fig:mapping}
\end{figure}

To better ground the framework introduced in \Cref{properties-of-human-centred-ai}, this section compares it with existing similar frameworks available in the literature. More precisely, three sets of principles, are taken as reference for the comparison \citep{Fjeld2020Principled,Maslej2024Artificial,Shneiderman2022HumanCentered}.
The comparison among them and the proposed human-centered \ac{SAI} principles is shown in \Cref{fig:mapping}.

In 2020, the Berkman Klein Centre published a report to inform policymakers, researchers, and industry stakeholders about the complex ethical challenges posed by \ac{AI} \citep{Fjeld2020Principled}. It promotes the development and deployment of \ac{AI} technologies that respect humans' fundamental rights, values, and societal norms. The report maps 36 documents illustrating frameworks, guidelines, and principles proposed by governments, international organizations, academic institutions, and industries. It delineates eight principles to guide the creation of \ac{AI}-based systems: \emph{Transparency and Explainability}, \emph{Fairness and Non-Discrimination}, \emph{Safety and Security}, \emph{Accountability}, \emph{Promotion of Human Values}, \emph{Privacy}, \emph{Professional Responsibility}, and \emph{Human Control of Technology} \citep{Fjeld2020Principled}.

The principles differ from the ones identified in this report in structure, dependencies, and subdimensions. From the \ac{SLR}, a new underlying hierarchical structure emerged; for example, we consider \emph{Safety}, \emph{Security}, and \emph{Privacy} on the same level and part of the \emph{Protection} principle, whereas the report views Privacy as a separate concept. As opposed to the Berkman Klein Centre's Report, the principle of \emph{Transparency} in our proposal stands at a higher level than \emph{Explainability} and \emph{Interpretability}, which indicates that an \ac{AI} system exhibits transparent behavior if it presents models that provide appropriate and understandable explanations that allow humans to interpret their outputs correctly. Similar considerations hold for the principle of \emph{Fairness}: from the content analysis we performed, a \emph{fair} behavior of an \ac{AI} model is achieved through a non-discriminatory approach, respecting human rights and spreading legitimate information. In this context, \emph{Promotion of Human Values} was not identified as a separate or standalone principle since it is the leitmotif of the framework: each principle aims at ensuring that human values, rights, and ethics are preserved and reflected in the interaction with products that feature \ac{AI}. As the law in the European Union becomes more involved in the creation of \ac{AI} systems, designers, developers and operators are taking on more responsibility, which means that they must comply with regulations ``by design''. Concerning \emph{Human Control of Technology}, it plays a crucial role in our review, and it was redefined as part of a greater principle, called \emph{Automation Level}, in which we indicate that \ac{AI} systems exhibit a spectrum of automated behavior with respect to human intervention, ranging from guaranteeing oversight to allowing complete control \citep{Fjeld2020Principled}.

Integrating \ac{HCI} in the design and development of AI is a relatively new area of research. Ben Shneiderman, one of the pioneers of \ac{HCAI}, focuses on three main properties--\textit{Trustworthiness}, \textit{Safety}, and \textit{Reliability}--as the key to reach other principles and satisfy the desiderata of \ac{AI} systems. Shneiderman stresses the role of a proper balance between automation and control to empower humans \citep{Shneiderman2022HumanCentered}. With respect to this view of \ac{HCAI}, the framework proposed in this manuscript splits and incorporates the three values in different principles. For example, we argue that a system is trustworthy if it follows multiple principles. Similarly, a \emph{safe} system is a system that protects users, chooses the right automation level for the task at hand, and behaves fairly \citep{ISO20199241210}. Finally, \emph{reliability} of a system is a property that is commonly known and used in Software Engineering, and requires the intelligibility of the system's operations, guaranteed by our framework's principles of transparency and lawfulness while ensuring that the system behaves in accordance with the purposes of the interaction \citep{Siegel2024Media}.

Concurrently with the release of the \ac{AI} Act in the \ac{EU}, in 2024, Stanford University published the Artificial Intelligence Index Report, which reports the current trends, tracks, and advancements in the context of \ac{AI}. This report presents ten key takeaways that highlight how \ac{AI} cannot fully replace humans in terms of dimensions like reasoning, common sense, and empathy. Industries continue to excel in \ac{AI} research, with the United States leading in developing top models, though these advancements come with rising costs not only in terms of resources but also in ethics and law. In addition, the scientific progress in the context of Generative \ac{AI} has proven to be a boost in productivity, but there is still a notable lack when it comes to evaluating the responsibility of \ac{AI} models, which is addressed in the report in the \emph{Responsible \ac{AI}} chapter. It provides its definition and dimensions while stressing that the ethical misuse of \ac{AI}-based systems and their fast worldwide spread is contributing to a rise in the number of incidents \citep{Maslej2024Artificial}.

Although the report focuses on \ac{AI}'s geopolitical and economic trends, some of its takeaways reflect the principles that emerged from our \ac{SLR}. For instance, Automation Level is significantly present in the first and seventh takeaways being, respectively ``\ac{AI} beats humans on some tasks, but not all'' and ``The data is in: \ac{AI} makes workers more productive and leads to higher quality work''; as humans possess characteristics that cannot be fully reproduced by \ac{AI}, therefore their oversight and control becomes necessary. The ninth and tenth takeaways ``The number of \ac{AI} regulations {[}in the United States{]} sharply increases'' and ``People across the globe are more cognizant of \ac{AI}'s potential impact---and more nervous'' relate to the Lawfulness principle, as it highlights how relevant the legal component influences the design and development of \ac{AI} systems. The fifth takeaway reflects the property of Robustness, one of the meta-codes of our research; it states that ``robust and standardized evaluations for \ac{LLM} responsibility are seriously lacking'' \citep{Maslej2024AI}.

The four key dimensions that characterize Responsible \ac{AI} are: \textit{Privacy and Data Governance}, \textit{Transparency and Explainability}, \textit{Security and Safety}, and \textit{Fairness}. Referring to our framework, data governance is an intrinsic aspect of Protection since it refers to the establishment of policies to guarantee the security, quality, and safe use of data, while \emph{Explainability} is a dimension of the \emph{Transparency} principle.

The principles that emerged from our \ac{SLR} show how many of the concepts at the basis of the current literature are still relevant, but they should be enhanced or tweaked to comply with the new requirements. Our research highlights how the leitmotif of the research in the era of \ac{AI} is the importance of humans; although previous literature already undertook this approach, including it in a legal and obligatory framework inevitably changes the perspective of researchers and companies.

\subsection{Uncovering Current Trends and Challenges}\label{uncovering-current-trends-and-challenges}

Throughout the literature review, some interesting trends and relevant challenges emerged that are worth reporting. This section briefly discusses them, highlighting open questions that may guide future research for the development of \ac{SAI} systems that comply with the \ac{AI} Act.

\begin{figure}[b!]
    \centering
    \includegraphics[width=.7\linewidth]{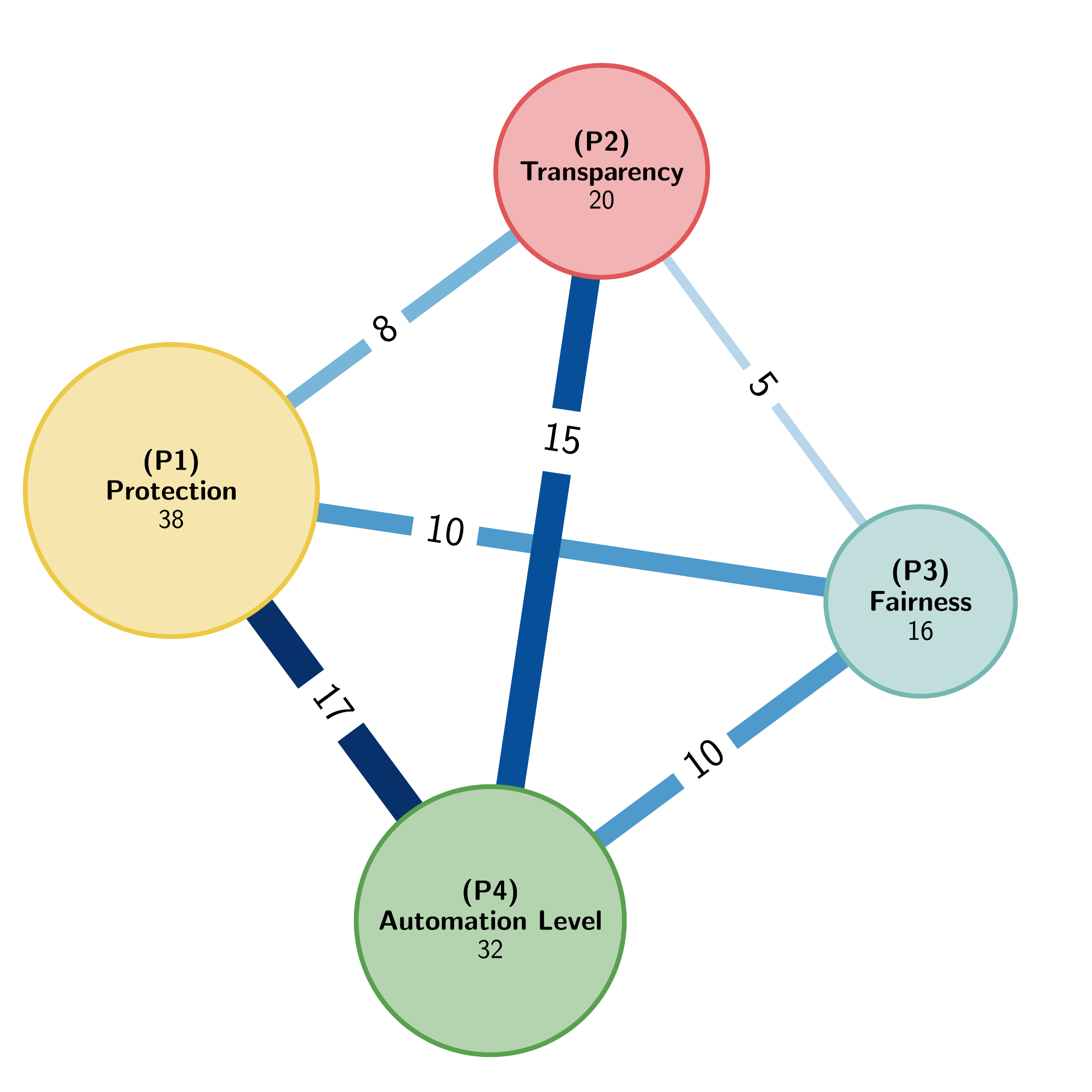}
    \caption{The various identified principles and their relationships. Each node size represents the number of articles classified in each principle (the nodes' labels report the number). The edge size and colour represent the number of articles classified in multiple principles (the edges' labels report the number).}
    \label{fig:principles-relationships}
\end{figure}

\subsubsection{Research Trends}\label{research-trends}
The research trends that emerged from this \ac{SLR} are represented in \Cref{fig:principles-relationships}, which highlight the extent to which principles are considered in the literature and how they are related to each other. It is important to underline that researchers seem to be interested in investigating aspects of the identified principles from a legal perspective considering their impact on legal implications regarding the \ac{AI} Act, the new set of soft laws that promises to bring breaking changes in the way \ac{AI} is designed and developed.

\paragraph{Trend 1  -- Increasing focus on Transparency}\label{trend-1-increasing-focus-on-transparency}
Analysing the review results, there appears to be a lack of emphasis on \ac{AI} algorithms' transparency among researchers in articles published before 2024. This concern is overcome in subsequent works, which can be attributed to the recent adoption of the \ac{AI} Act, which prioritizes legal considerations over technical ones in the way humans can understand the \ac{AI} behaviors. Furthermore, while the \ac{AI} Act offers guidelines for developing compliant \ac{AI} systems, it does not provide specific technical instructions. Consequently, researchers are working on understanding how to implement it algorithmically.

\paragraph{Trend 2 -- Significant Connection among Fairness with Protection and Automation Level}\label{trend-2-connection-among-protection-fairness-and-automation}
Researchers seem to be focused on investigating not only the legal considerations but also on crafting methods to safeguard users while ensuring non-discriminatory and ethical \ac{AI} practices. Fairness exhibits equal connection with Protection and Automation Level, as following practices that preserve human rights translates in \ac{AI} systems that handle sensitive personal data in the proper way while ensuring that individuals can exercise control over their behaviour.

\paragraph{Trend 3 -- Weak connection among Transparency and Fairness}\label{trend-3-weak-connection-among-transparency-and-fairness}
In the current scenario, algorithmic transparency and fairness are not always investigated together. In fact, from the review emerged that Transparency and Fairness are weakly correlated; while both are crucial for developing an ethical \ac{AI}, they often require different approaches and considerations.

\paragraph{Trend 4 -- Strong connection among Automation Level, Transparency and Protection}\label{trend-4-strong-connection-automationLevel-transparency-protection}
The level of automation within a system is closely tied to its transparency and the associated protective measures to enhance human oversight and system protection. Transparent systems allow humans to understand their behavior, enabling better supervision, effective intervention, and the identification of vulnerabilities. This can reduce the occurrence of unexpected events potentially harming humans, and transparency improves safety by balancing automation and augmentation.

\subsubsection{Research Challenges}\label{research-challenges}
This \ac{SLR} revealed some of the challenges of the current landscape of \ac{AI} research, ranging from the lack of technical solutions for the new legal constraints and requirements to a missing shared standpoint among researchers. Another key issue remains Trustworthiness, which impacts human-\ac{AI} relationship but has the potential of damaging the decision-making process.

\paragraph{Challenge 1 -- Lack of technical design solutions}\label{challenge-1-lack-of-technical-solutions}
Most of the research does not conduct studies or experiments aiming at proposing new technical solutions, thus favouring a more general discussion. These results should not be surprising. In fact, the \ac{AI} Act discusses the need for safe, ethical, fair, and trustworthy \ac{AI} systems. However, no practical indications on how to design and develop such \ac{AI} systems are provided. This implies that scholars and practitioners lack technical guidance to ensure the compliance of new \ac{AI} solutions.

\paragraph{Challenge 2 -- Lack of standardized evaluation methods}\label{challenge-2-lack-of-standardized-evaluation-methods}
The \ac{AI} Act remains a legal framework, providing theoretical and conceptual indications concerning the creation of \ac{AI} systems. Although the \ac{EU} also proposes a tool for compliance checking with the regulation, there is still the need for a standardized approach in the evaluation and assessment of the properties that characterize \ac{AI} systems. This can translate to quantitative or qualitative methods that can objectively identify the strengths and weaknesses of the system.

\paragraph{Challenge 3 -- Lack of a common view}\label{challenge-3-lack-of-a-common-view}
Most of the analyzed papers do not suggest a common view of the proposed solutions (e.g., basic definitions, guidelines, frameworks, etc.) It emerged that a standardized approach is still lacking, and there are no methodological common approaches that can be employed to design and develop \ac{AI} Act-compliant systems. The research community seems fragmented, focusing on different aspects of the matter, even exhibiting opposite standpoints.

\paragraph{Challenge 4 -- Is there anything beyond trust?}\label{challenge-4-anything-beyond-trust}
Given the previous legal documents that contributed to laying its groundwork, the \ac{AI} Act heavily relies on Trustworthy \ac{AI}, which can potentially influence the perspective of the articles related to the regulation. As a result, many research works frequently highlight trustworthiness as the umbrella property of \ac{AI}, which inevitably impacts the generalization of the concepts that emerged from this literature review. Although trustworthiness is an important component of the interaction between humans and \ac{AI}, it must be carefully evaluated and balanced with other factors.

\paragraph{Challenge 5 -- Inconsistencies in terminology}\label{challenge-5-inconsistencies-in-terminology}
\ac{AI} impacts countless aspects of modern society, implying that governmental bodies, researchers, and end users must be aligned in the use of terms and concepts that revolve around the design, development, deployment, and use of \ac{AI} systems. From this \ac{SLR}, it emerged that there are words that possess different meanings depending on the expertise of those who use them. For example, the term \textit{Transparency} can refer either to the transparent use and storage of data in case of the \ac{GDPR}, or to the extent to which an \ac{AI} model is transparent to humans in the case of the \ac{AI} Act. There is also a lack of clarity concerning the differences between the \textit{human-centred} and the \textit{human-centric} approaches, which are mistakenly often used interchangeably. This issue highlights the need for more uniformity in digital literacy for an aligned and aware society.

\paragraph{Challenge 6 -- Underexplored impact of human factors}\label{challenge-6-underexplored-impact-of-human-factors}
The increasing adoption of the human-centred approach to creating \ac{AI} systems highlights the need for a more in-depth study of the human factors that can influence its relationship with humans. This \ac{SLR} revealed the current research concerning the \ac{AI} Act mainly revolves around Law and Computer Science, leaving out some relevant aspects concerning the psychological and behavioural factors that can impact the establishment of a symbiotic relationship between humans and \ac{AI}. This aspect needs to be further investigated to ensure the seamless integration of these systems in our daily lives and guarantee the augmentation of humans instead of their replacement.

\subsection{Limitations}\label{limitations-srq3}

In general, several threats to validity can affect the results of a \ac{SLR}. In the following, we report how we mitigated the most critical ones.

\paragraph{Selection bias}\label{selection-bias}
This happens when the research papers selected for the review do not represent all the studies conducted on the topic. In fact, the personal biases of the reviewers can influence the selection and interpretation of papers. This was mitigated by deeply analysing the selected papers to check their compliance with the objectives of the \ac{SLR} and by involving multiple reviewers who independently assessed them.

\paragraph{Publication bias}\label{publication-bias}
This occurs when studies that show statistically significant results are more likely to be published than studies that do not. This aspect did not occur while performing this literature review since most of the selected papers concern the legal field.

\paragraph{Time lag bias}\label{time-lag-bias}
This arises when not all the relevant works are included in the \ac{SLR} due to their publication after the review was conducted. In this case, the work was performed one month before its submission.

\paragraph{Publication quality}\label{publication-quality}
This emerges when poor quality works are considered in the \ac{SLR}. To mitigate this aspect, inclusion and exclusion criteria considering the quality of the publication venue were defined leading to a manual evaluation of publications that appeared in venues of lower quality.

\section{Conclusions}
The spreading of \ac{AI} in many contexts of our daily lives leads to the necessity of understanding its applications and potential improvements along with its limitations. Due to the rapid advancement of \ac{AI}, ethical and legal aspects have to be carefully addressed when creating these systems with the objective to considering humans in all of their dimensions, and not as mere users not undermining their capabilities and avoiding any form of discrimination \citep{ISO20199241210, Stahl2023Unfair}.
Both academia and government are identifying solutions by promoting a new perspective called \ac{HCAI} which represents the need to adopt a human-centred approach when creating \ac{AI} systems. \ac{HCAI} is the foundation to create \ac{SAI} systems that foster a human-\ac{AI} symbiotic relationship empowering humans rather than replacing them \citep{Desolda2024Human}.
To safeguard humans during the interaction process and the environment surrounding them, the \ac{EU} enacted the \ac{AI} Act regulating their usage employing a risk-based approach \citep{EuropeanParliament2024Regulation}.

This article aims to present the principles, resulted from a \ac{SLR}, that can guide practitioners to create \ac{SAI} systems while being compliant with the \ac{AI} Act considering humans as the core of the development process. Confirming what \cite{Desolda2024Humancentered} argued, our \ac{SLR} highlights that  building \ac{SAI} systems requires a multidisciplinary and human-centred approach, focusing on supporting humans rather than replacing them \citep{Shneiderman2022HumanCentered}. This implies ensuring \ac{AI}'s ethical and lawful behaviour as one of the main crucial aspects to integrate in the interaction between humans and \ac{AI}, highlighting that both parties should improve over time, learning from each other \citep{Desolda2024Humancentered}. Although there are technical challenges that designers and developers could encounter when creating \ac{SAI} systems, ethical and philosophical issues must be inevitably addressed during the process \citep{lisiAcceptabilitySymbioticArtificial2024}.

This research has proposed a framework for building \ac{SAI} systems that comply with the \ac{AI} Act, composed of four principles--- \emph{Transparency}, \emph{Fairness}, \emph{Automation Level}, and \emph{Protection}---and three properties---\emph{Trustworthiness}, \emph{Robustness}, and \emph{Sustainability}. Both principles and properties foster symbiotic relationship, each to different extents, while guiding practitioners in creating legally-compliant and ethical systems. The results of this study lay the groundwork for the creation of a holistic framework, composed of guidelines and success criteria, that can provide them with more complete and concrete support through both design and evaluation methods, practices, and patterns.

\bibliography{references}
\defaultbibliographystyle{sn-apacite}
\defaultbibliography{references}







\section*{Statements and Declarations}



\subsection*{Funding}

This research is partially supported by:

\begin{itemize}
    \item the co-funding of the European Union - Next Generation EU: NRRP Initiative, Mission 4, Component 2, Investment 1.3 – Partnerships extended to universities, research centers, companies and research D.D. MUR n. 341 del 15.03.2022 – Next Generation EU (PE0000013 – “Future Artificial Intelligence Research – FAIR” - CUP: H97G22000210007);
    \item the Italian Ministry of University and Research (MUR) and by the European Union - NextGenerationEU, Mission 4, Component 2, Investment 1.1, under grant PRIN 2022 PNRR ``PROTECT: imPROving ciTizEn inClusiveness Through Conversational AI'' (Grant P2022JJPBY) — CUP: H53D23008150001.
    \item the Italian Ministry of University and Research (MUR) under grant PRIN 2022 “DevProDev: Profiling Software Developers for Developer-Centered Recommender Systems” — CUP: H53D23003620006;
\end{itemize}

The research of Andrea Esposito is funded by a Ph.D. fellowship within the framework of the Italian ``D.M. n. 352, April 9, 2022''- under the National Recovery and Resilience Plan, Mission 4, Component 2, Investment 3.3 -- Ph.D. Project ``Human-Centred Artificial Intelligence (HCAI) techniques for supporting end users interacting with AI systems,'' co-supported by ``Eusoft S.r.l.'' (CUP H91I22000410007).

The research of Antonio Curci and Miriana Calvano is supported by the co-funding of the European Union - Next Generation EU: NRRP Initiative, Mission 4, Component 2, Investment 1.3 -- Partnerships extended to universities, research centers, companies, and research D.D. MUR n. 341 del 15.03.2022 -- Next Generation EU (PE0000013 -- ``Future Artificial Intelligence Research -- FAIR'' - CUP: H97G22000210007).

\subsection*{Author Contributions Statement}

\textbf{Conceptualization:} Miriana Calvano, Antonio Curci, Rosa Lanzilotti, Antonio Piccinno;
\textbf{Data curation:} Miriana Calvano, Antonio Curci, Andrea Esposito;
\textbf{Formal analysis:} Miriana Calvano, Antonio Curci;
\textbf{Funding acquisition:} Rosa Lanzilotti;
\textbf{Investigation:} Miriana Calvano, Antonio Curci;
\textbf{Methodology:} Miriana Calvano, Antonio Curci, Andrea Esposito, Rosa Lanzilotti, Antonio Piccinno;
\textbf{Project administration:} Rosa Lanzilotti;
\textbf{Supervision:} Giuseppe Desolda, Andrea Esposito, Rosa Lanzilotti, Antonio Piccinno;
\textbf{Validation:} Andrea Esposito;
\textbf{Visualization:} Andrea Esposito;
\textbf{Writing -- original draft:} Miriana Calvano, Antonio Curci, Andrea Esposito;
\textbf{Writing -- review \& editing:} Miriana Calvano, Antonio Curci, Giuseppe Desolda, Andrea Esposito, Rosa Lanzilotti, Antonio Piccinno.

\subsection*{Competing Interests}
The authors declare no competing interests.

\subsection*{Data Availability Statement}

No new data were created or analyzed during this study. Data sharing is not applicable to this article. All relevant literature records used within this literature review are cited within the body of the manuscript and can be found in the References section.

\newpage
\appendix
\section{Appendix}
\subsection{Classification of the Selected Manuscripts}\label{classification-of-the-selected-manuscripts}
\Cref{tab:map-articles-principles,tab:map-principles-properties} present the classification of the selected articles, providing two different views. \Cref{tab:map-articles-principles} presents, for each selected article, the various dimensions that are associated with all the principles. \Cref{tab:map-principles-properties} provides an overview of which article discusses a property in the context of each principle.

\begin{landscape}
\begin{longtblr}[
  caption = {Mapping between the articles and the principles},
  label = {tab:map-articles-principles}
]{
  colspec = {m{3cm}ccccc},
  rowhead = 1,
  rows = {font=\footnotesize}
}
\hline
\textbf{Ref.} & \textbf{Protection (P1)} & \textbf{Transparency (P2)} & \textbf{Fairness (P3)} & \textbf{Automation Level (P4)} \\ \hline
\textbf{\cite{Helberger2023ChatGPT}} & \begin{tabular}[c]{@{}c@{}}Safety\end{tabular} &  &  &  \\
\textbf{\cite{Lazcoz2023Humans}} &  & \begin{tabular}[c]{@{}c@{}}Interpretability\end{tabular} &  & \begin{tabular}[c]{@{}c@{}}Human-In-The-Loop;\\ Human-On-The-Loop\end{tabular} \\
\textbf{\cite{Malgieri2024Licensing}} &  & \begin{tabular}[c]{@{}c@{}}Explainability\end{tabular} &  &  \\
\textbf{\cite{Varosanec2022Path}} &  & \begin{tabular}[c]{@{}c@{}}Explainability\end{tabular} &  & \begin{tabular}[c]{@{}c@{}}Human-On-The-Loop\end{tabular} \\
\textbf{\cite{Urquhart2022Legal}} &  & \begin{tabular}[c]{@{}c@{}}Explainability;\\ Interpretability\end{tabular} &  & \begin{tabular}[c]{@{}c@{}}Human-On-The-Loop\end{tabular} \\
\textbf{\cite{Kieseberg2023Controllable}} & \begin{tabular}[c]{@{}c@{}}Security\end{tabular} & \begin{tabular}[c]{@{}c@{}}Explainability\end{tabular} &  & \begin{tabular}[c]{@{}c@{}}Human-In-The-Loop\end{tabular} \\
\textbf{\cite{Tartaro2023European}} & \begin{tabular}[c]{@{}c@{}}Safety\end{tabular} &  &  & \begin{tabular}[c]{@{}c@{}}Human-On-The-Loop\end{tabular} \\
\textbf{\cite{Enqvist2023Human}} &  &  &  & \begin{tabular}[c]{@{}c@{}}Human-On-The-Loop\end{tabular} \\
\textbf{\cite{Neuwirth2023Prohibited}} & \begin{tabular}[c]{@{}c@{}}Safety\end{tabular} &  &  &  \\
\textbf{\cite{Gyevnar2023Bridging}} &  & \begin{tabular}[c]{@{}c@{}}Explainability\end{tabular} &  &  \\
\textbf{\cite{Stuurman2022Regulating}} & \begin{tabular}[c]{@{}c@{}}Privacy\end{tabular} &  & \begin{tabular}[c]{@{}c@{}}Non-Discrimination\end{tabular} &  \\
\textbf{\cite{Mueck2023Upcoming}} & \begin{tabular}[c]{@{}c@{}}Privacy;\\ Security\end{tabular} &  &  &  \\
\textbf{\cite{Hacker2023European}} & \begin{tabular}[c]{@{}c@{}}Safety\end{tabular} &  &  & \begin{tabular}[c]{@{}c@{}}Human-On-The-Loop\end{tabular} \\
\textbf{\cite{Porlezza2023Promoting}} &  &  & \begin{tabular}[c]{@{}c@{}}Rightful Information\end{tabular} &  \\
\textbf{\cite{Chamberlain2023RiskBased}} &  &  &  & \begin{tabular}[c]{@{}c@{}}Human-In-The-Loop\end{tabular} \\
\textbf{\cite{Hupont2022Landscape}} & \begin{tabular}[c]{@{}c@{}}Privacy;\\ Safety;\\ Security\end{tabular} &  & \begin{tabular}[c]{@{}c@{}}Non-Discrimination\end{tabular} & \begin{tabular}[c]{@{}c@{}}Human-In-The-Loop;\\ Human-On-The-Loop\end{tabular} \\
\textbf{\cite{Mylly2023Transparent}} & \begin{tabular}[c]{@{}c@{}}Safety\end{tabular} &  &  & \begin{tabular}[c]{@{}c@{}}Human-In-The-Loop\end{tabular} \\
\textbf{\cite{Anamaria2023Artificial}} & \begin{tabular}[c]{@{}c@{}}Privacy\end{tabular} &  &  &  \\
\textbf{\cite{Hupont2023Documenting}} & \begin{tabular}[c]{@{}c@{}}Privacy;\\ Security\end{tabular} &  &  &  \\
\textbf{\cite{Pagallo2022Environmental}} & \begin{tabular}[c]{@{}c@{}}Safety\end{tabular} &  &  &  \\
\textbf{\cite{Morozovaite2023Hypernudging}} & \begin{tabular}[c]{@{}c@{}}Privacy\end{tabular} &  &  &  \\
\textbf{\cite{Laux2023Institutionalised}} &  & \begin{tabular}[c]{@{}c@{}}Explainability\end{tabular} &  & \begin{tabular}[c]{@{}c@{}}Human-On-The-Loop\end{tabular} \\
\textbf{\cite{Mazur2024Embedding}} &  &  &  & \begin{tabular}[c]{@{}c@{}}Human-In-The-Loop\end{tabular} \\
\textbf{\cite{Herbosch2024Fraud}} &  &  & \begin{tabular}[c]{@{}c@{}}Rightful Information\end{tabular} & \begin{tabular}[c]{@{}c@{}}Human-On-The-Loop\end{tabular} \\
\textbf{\cite{RomeroMoreno2024Generative}} & \begin{tabular}[c]{@{}c@{}}Privacy;\\ Safety\end{tabular} &  &  &  \\
\textbf{\cite{Sposini2024Neuromarketing}} &  &  & \begin{tabular}[c]{@{}c@{}}Rightful Information\end{tabular} &  \\
\textbf{\cite{Yordanova2024Regulating}} & \begin{tabular}[c]{@{}c@{}}Privacy\end{tabular} &  &  &  \\
\textbf{\cite{Biewer2024Software}} &  &  &  & \begin{tabular}[c]{@{}c@{}}Human-On-The-Loop\end{tabular} \\
\textbf{\cite{CoveloDeAbreu2024Artificial}} &  &  &  & \begin{tabular}[c]{@{}c@{}}Human-In-The-Loop;\\ Human-On-The-Loop\end{tabular} \\
\textbf{\cite{Stettinger2024Trustworthiness}} &  & \begin{tabular}[c]{@{}c@{}}Explainability;\\ Interpretability\end{tabular} & \begin{tabular}[c]{@{}c@{}}Non-Discrimination\end{tabular} & \begin{tabular}[c]{@{}c@{}}Human-In-The-Loop;\\ Human-On-The-Loop\end{tabular} \\
\textbf{\cite{Almada2024Brussels}} & \begin{tabular}[c]{@{}c@{}}Privacy;\\ Safety;\\ Security\end{tabular} &  &  &  \\
\textbf{\cite{Aseeva2023Liable}} & \begin{tabular}[c]{@{}c@{}}Privacy;\\ Safety\end{tabular} &  &  &  \\
\textbf{\cite{Helberger2024FutureNewsCorp}} & \begin{tabular}[c]{@{}c@{}}Safety;\\ Security\end{tabular} &  &  & \begin{tabular}[c]{@{}c@{}}Human-On-The-Loop\end{tabular} \\
\textbf{\cite{Hirvonen2023Just}} & \begin{tabular}[c]{@{}c@{}}Safety\end{tabular} &  &  &  \\
\textbf{\cite{Hupont2024Use}} & \begin{tabular}[c]{@{}c@{}}Safety\end{tabular} &  &  &  \\
\textbf{\cite{Laux2024Trustworthy}} &  & \begin{tabular}[c]{@{}c@{}}Explainability\end{tabular} & \begin{tabular}[c]{@{}c@{}}Non-Discrimination\end{tabular} & \begin{tabular}[c]{@{}c@{}}Human-In-The-Loop\end{tabular} \\
\textbf{\cite{Migliorini2024More}} & \begin{tabular}[c]{@{}c@{}}Privacy\end{tabular} &  & \begin{tabular}[c]{@{}c@{}}Rightful Information\end{tabular} &  \\
\textbf{\cite{Nikiforov2024Groups}} & \begin{tabular}[c]{@{}c@{}}Privacy\end{tabular} &  &  &  \\
\textbf{\cite{Olsen2024Right}} &  & \begin{tabular}[c]{@{}c@{}}Explainability;\\ Interpretability\end{tabular} &  &  \\
\textbf{\cite{Pavlidis2024Unlocking}} & \begin{tabular}[c]{@{}c@{}}Privacy;\\ Security\end{tabular} & \begin{tabular}[c]{@{}c@{}}Explainability;\\ Interpretability\end{tabular} &  & \begin{tabular}[c]{@{}c@{}}Human-On-The-Loop\end{tabular} \\
\textbf{\cite{Siegel2024Media}} & \begin{tabular}[c]{@{}c@{}}Security\end{tabular} & \begin{tabular}[c]{@{}c@{}}Explainability\end{tabular} & \begin{tabular}[c]{@{}c@{}}Rightful Information\end{tabular} & \begin{tabular}[c]{@{}c@{}}Human-In-The-Loop;\\ Human-On-The-Loop\end{tabular} \\
\textbf{\cite{Tallberg2024AI}} & \begin{tabular}[c]{@{}c@{}}Safety;\\ Security\end{tabular} & \begin{tabular}[c]{@{}c@{}}Explainability\end{tabular} &  & \begin{tabular}[c]{@{}c@{}}Human-On-The-Loop\end{tabular} \\
\textbf{\cite{Wagner2024Navigating}} & \begin{tabular}[c]{@{}c@{}}Safety;\\ Security\end{tabular} & \begin{tabular}[c]{@{}c@{}}Explainability\end{tabular} &  &  \\
\textbf{\cite{Worsdorfer2023Mitigating}} & \begin{tabular}[c]{@{}c@{}}Privacy;\\ Safety\end{tabular} &  & \begin{tabular}[c]{@{}c@{}}Non-Discrimination;\\ Rightful Information\end{tabular} & \begin{tabular}[c]{@{}c@{}}Human-On-The-Loop\end{tabular} \\
\textbf{\cite{Kattnig2024Assessing}} & \begin{tabular}[c]{@{}c@{}}Privacy\end{tabular} & \begin{tabular}[c]{@{}c@{}}Explainability\end{tabular} & \begin{tabular}[c]{@{}c@{}}Non-Discrimination\end{tabular} & \begin{tabular}[c]{@{}c@{}}Human-In-The-Loop\end{tabular} \\
\textbf{\cite{Montagnani2024EU}} & \begin{tabular}[c]{@{}c@{}}Privacy;\\ Safety\end{tabular} &  &  & \begin{tabular}[c]{@{}c@{}}Human-On-The-Loop\end{tabular} \\
\textbf{\cite{Laux2024Three}} & \begin{tabular}[c]{@{}c@{}}Privacy;\\ Safety\end{tabular} &  & \begin{tabular}[c]{@{}c@{}}Non-Discrimination;\\ Rightful Information\end{tabular} &  \\
\textbf{\cite{BoteroArcila2024AI}} & \begin{tabular}[c]{@{}c@{}}Safety\end{tabular} &  & \begin{tabular}[c]{@{}c@{}}Non-Discrimination;\\ Rightful Information\end{tabular} & \begin{tabular}[c]{@{}c@{}}Human-In-The-Loop;\\ Human-On-The-Loop\end{tabular} \\
\textbf{\cite{Baumgartner2024AIDriven}} & \begin{tabular}[c]{@{}c@{}}Privacy;\\ Security\end{tabular} & \begin{tabular}[c]{@{}c@{}}Explainability;\\ Interpretability\end{tabular} &  & \begin{tabular}[c]{@{}c@{}}Human-In-The-Loop\end{tabular} \\
\textbf{\cite{VanKolfschooten2024EU}} & \begin{tabular}[c]{@{}c@{}}Safety\end{tabular} &  &  & \begin{tabular}[c]{@{}c@{}}Human-On-The-Loop\end{tabular} \\
\textbf{\cite{Lee2024Transformative}} & \begin{tabular}[c]{@{}c@{}}Safety\end{tabular} &  &  & \begin{tabular}[c]{@{}c@{}}Human-On-The-Loop\end{tabular} \\
\textbf{\cite{CanteroGamito2024Artificial}} &  & \begin{tabular}[c]{@{}c@{}}Explainability;\\ Interpretability\end{tabular} &  & \begin{tabular}[c]{@{}c@{}}Human-On-The-Loop\end{tabular} \\
\textbf{\cite{Peeters2024Editorial}} &  & \begin{tabular}[c]{@{}c@{}}Explainability\end{tabular} &  & \begin{tabular}[c]{@{}c@{}}Human-In-The-Loop;\\ Human-On-The-Loop\end{tabular} \\
\textbf{\cite{Nisevic2024Explainable}} & \begin{tabular}[c]{@{}c@{}}Privacy\end{tabular} & \begin{tabular}[c]{@{}c@{}}Explainability\end{tabular} &  &  \\
\textbf{\cite{Novelli2024Robust}} & \begin{tabular}[c]{@{}c@{}}Privacy;\\ Safety\end{tabular} &  &  &  \\
\textbf{\cite{Cupac2024Regulate}} & \begin{tabular}[c]{@{}c@{}}Privacy;\\ Safety\end{tabular} &  & \begin{tabular}[c]{@{}c@{}}Non-Discrimination;\\ Rightful Information\end{tabular} &  \\
\textbf{\cite{Papadakis2024Explainable}} &  & \begin{tabular}[c]{@{}c@{}}Explainability;\\ Interpretability\end{tabular} & \begin{tabular}[c]{@{}c@{}}Non-Discrimination\end{tabular} & \begin{tabular}[c]{@{}c@{}}Human-In-The-Loop\end{tabular} \\
\textbf{\cite{Gornet2024European}} & \begin{tabular}[c]{@{}c@{}}Safety\end{tabular} &  & \begin{tabular}[c]{@{}c@{}}Non-Discrimination\end{tabular} & \begin{tabular}[c]{@{}c@{}}Human-On-The-Loop\end{tabular} \\
\hline
\end{longtblr}
\end{landscape}

\begin{landscape}
\begin{longtblr}
[
  caption = {Mapping the articles between principles and properties},
  label = {tab:map-principles-properties}
]{
  colspec = {m{2cm}XXX},
  rowhead = 1,
  rows = {font=\footnotesize}
}
    \textbf{ } & \textbf{Trustworthiness} & \textbf{Robustness} & \textbf{Sustainability} \\ \hline 
        \textbf{Protection} & \cite{Kieseberg2023Controllable, Tartaro2023European, Neuwirth2023Prohibited, Stuurman2022Regulating, Hacker2023European, Hupont2022Landscape, Mylly2023Transparent, Anamaria2023Artificial, Hupont2023Documenting, Pagallo2022Environmental, Almada2024Brussels, Helberger2024FutureNewsCorp, Hupont2024Use, Nikiforov2024Groups, Siegel2024Media, Wagner2024Navigating, Worsdorfer2023Mitigating, Kattnig2024Assessing, Montagnani2024EU, Laux2024Three, BoteroArcila2024AI, Baumgartner2024AIDriven, Lee2024Transformative, CanteroGamito2024Artificial, Peeters2024Editorial, Nisevic2024Explainable, Cupac2024Regulate, Papadakis2024Explainable, Gornet2024European} & \cite{Kieseberg2023Controllable, Hupont2022Landscape, Hupont2023Documenting, Nikiforov2024Groups, Pavlidis2024Unlocking, Wagner2024Navigating, Baumgartner2024AIDriven, Lee2024Transformative, Tallberg2024AI, Novelli2024Robust} & \\
        \textbf{Fairness} & \cite{Stuurman2022Regulating, Siegel2024Media, Worsdorfer2023Mitigating, Kattnig2024Assessing, Laux2024Three, BoteroArcila2024AI, Cupac2024Regulate, Papadakis2024Explainable, Gornet2024European, Stettinger2024Trustworthiness, Laux2024Trustworthy} &  & \cite{Siegel2024Media, Worsdorfer2023Mitigating, Laux2024Three} \\   
        \textbf{Automation Level} & \cite{Urquhart2022Legal,Varosanec2022Path, Kieseberg2023Controllable, Tartaro2023European, Hacker2023European, Hupont2022Landscape, Mylly2023Transparent, Helberger2024FutureNewsCorp, Siegel2024Media, Worsdorfer2023Mitigating, Kattnig2024Assessing, Montagnani2024EU, BoteroArcila2024AI, Baumgartner2024AIDriven, Lee2024Transformative, CanteroGamito2024Artificial, Peeters2024Editorial, Gornet2024European, Laux2023Institutionalised, Stettinger2024Trustworthiness, Laux2024Three, Herbosch2024Fraud, Enqvist2023Human, Biewer2024Software, CoveloDeAbreu2024Artificial, Helberger2023ChatGPT} &  & \\ \hline
\end{longtblr}
\end{landscape}


\end{document}